\documentclass[preprint,amsmath,amssymb,amssymb,aps,pra]{revtex4}
\usepackage{graphicx,url}
\usepackage[utf8]{inputenc}  
\usepackage[english]{babel}
\usepackage{physics}
\usepackage{float}
\usepackage{url}
\usepackage{soul}
\usepackage{graphicx}
\usepackage{dcolumn}
\usepackage{bm}
\usepackage{times}
\usepackage{mathrsfs}
\usepackage{color}
\usepackage{todonotes}
\usepackage{natbib}
\begin{document}
\title{Temperature-dependent optical constants of nanometer-thin flakes of Fe(Te,Se) superconductor in the visible and near-infrared regime}
\author{Aswini K. Pattanayak, Jagi Rout, and Pankaj K. Jha$\footnote{Correspondence and requests for materials should be requested to P.K.J (pkjha@syr.edu)}$}
\affiliation{Quantum Technology Laboratory $\langle \text{Q}|\text{T}|\text{L}\rangle$, Department of Electrical Engineering and Computer Science, Syracuse University, Syracuse, NY 13244, USA}
\begin{abstract}
Iron chalcogenides superconductors, such as Fe(Te,Se) have recently garnered significant attention due to their simple crystal structure with a relatively easy synthesis process, high-temperature superconductivity, intrinsic topological band structure, and an unconventional pairing of superconductivity with ferromagnetism. Here, we report the complex in-plane refractive index measurement of nanometer-thin Fe(Te,Se) flake exfoliated from a single crystal FeTe$_{\text{0.6}}$Se$_{\text{0.4}}$ for photon wavelengths from 450 to 1100 nm over a temperature range from 4 K to 295 K. The results were obtained by employing a two-Drude model for the dielectric function of Fe(Te,Se), a multiband superconductor, and fitting the absolute optical reflection spectra using the transfer matrix method. A high extinction coefficient in the visible to near-infrared range makes nanometer-thin Fe(Te,Se) flakes a promising material for photodetection applications.  
\end{abstract}
\date{\today}
\maketitle

\noindent \textbf{I. INTRODUCTION}

\vspace{2mm}
The discovery of iron-based superconductors (IBS) La[O$_{1-x}$F$_{x}$]FeAs in 2008~\cite{Kamihara2008}, with the superconducting critical transition temperature $T_{c}$ of 26 K, introduced a new family of high-temperature superconductors besides copper-based materials~\cite{Muller1986}. Soon, it was demonstrated that $T_{c}$ can be elevated to $\sim$ 50 K in similar compounds by replacing Lanthanum (La) with rare-earth elements Cerium (Ce)~\cite{Chen2008}, Neodymium (Nd)~\cite{Ren2008b}, Praseodymium (Pr)~\cite{Ren2008a} or Samarium (Sm)~\cite{Ren2008c}. Similar to cuprates, the superconductivity in IBS has an unconventional origin associated with spin fluctuations rather than the conventional electron-phonon pairing mechanism~\cite{Hirschfeld2011,Scalapino1995}. Within the family of IBS, layered iron-chalcogenides such as FeSe (so-called \textquotedblleft11-phase\textquotedblright) have served as a key material to understanding the microscopic physics of unconventional superconductivity in IBS owing to their simple crystal structure, relatively straightforward synthesis process, and lower toxicity stemming from the absence of the toxic heavy metal arsenic (As). Although for bulk FeSe single crystal at ambient pressure, the $T_{c} = $ 8K~\cite{Hsu2008} is relatively low for a high-temperature superconductor. By doping FeSe with tellurium (Te) i.e., FeTe$_{x}$Se$_{1-x}$ ($0\leq x\leq 1)$, hereafter referred to as Fe(Te,Se), $T_{c}$ can be increased due to expansion of the lattice parameters and reaches up to $\sim$15 K at a concentration of about  $x \sim$ 0.5-0.7~\cite{Yeh2008a, Fang2008, Yeh2008b}. However, a further increase in Te concentration in Fe(Te,Se) lowers its $T_c$ due to the strengthening of antiferromagnetic ordering. In the extreme limit, $x$ = 1, superconductivity vanishes, i.e., bulk FeTe is not superconducting. 

Recently, Fe(Te,Se) has stimulated great interest owing to its inherent unconventional pairing of superconductivity with ferromagnetism~\cite{Paglione2010, Rameau2019, McLaughlin2021}, observation of exotic high-temperature topological superconductivity~\cite{Zhang2018, Zhang2019}, and Majorana bound states~\cite{Wang2018, Kong2019, Machida2019}. Remarkably, analogous to its parent compound FeSe, monolayer Fe(Te,Se) on SrTiO$_{3}$ exhibits interfaced-enhanced $T_c$ of 40 K ~\cite{Li2015}. In the normal (non-superconducting) state, Fe(Te,Se) exhibits poor metallic behavior due to strong electron correlation~\cite{Sales2009,Pourret2011,Lanata2013}. The optical conductivity measurement indicates a significant reduction in the Drude weight due to electron correlations, providing evidence of poor metallic behavior. Furthermore, electronic structure calculations~\cite{X2} and angle-resolved photoemission spectroscopy~\cite{X3,X4,X5} reveal that Fe(Te,Se) is a multiband superconductor. The band structure of Fe(Te,Se) consists of the hole bands at the $\Gamma$ point and the electron bands at the M point \cite{R1,R2}, and its optical properties are best described by the two-Drude model for dielectric function~\cite{Homes2011, Pimenov2013, Perucchi2014, Homes2015}. One member of the Fe(Te,Se) family, FeTe$_{\text{0.6}}$Se$_{\text{0.4}}$, whose superconducting transition temperature reaches up to 21 K with Li-NH$_3$ intercalation~\cite{Li2017}, has garnered interest not only as a candidate for topological quantum matter~\cite{Machida2019,Lin2023}, but also developing van der Waals Josephson junctions~\cite{Miyazawa2021, Qiu2023}. The topological superconductivity in FeTe$_{\text{0.6}}$Se$_{\text{0.4}}$ supports the S = 1 spin-triplet state in a supercurrent, offering a platform for Ising-type spin-pairing superconductors~\cite{Ohnishi2021}.  Furthermore, FeTe$_{\text{0.6}}$Se$_{\text{0.4}}$ with high critical current densities under magnetic fields is advantageous for practical applications~\cite{Tanaka2017}. 

Although the Fe(Te,Se) family has been extensively studied for its magnetic properties, the optical properties of nanometer-thin films or flakes for photonic applications in visible to near-IR wavelengths over a wide range of temperatures remain relatively unexplored. An essential step toward potential photonics applications of FeTe$_{\text{0.6}}$Se$_{\text{0.4}}$ would be to determine its complex refractive index $\tilde{n} = n+ik$ over a wide range of temperatures and wavelengths. The complex refractive index $\tilde{n}$ of a superconductor offers crucial insights into how electromagnetic waves interact with the material, which is crucial for designing and optimizing devices for photonics and quantum photonics applications. For instance, the superconducting material's $\tilde{n}$ value is crucial for the cavity design for optimizing the photon detection efficiency of superconducting single-photon detectors~\cite{Rosfjord2006}.

In this work, we report the measurement of the complex-valued in-plane refractive index $\tilde{n}$ of nanometer-thin flake of Fe(Te,Se) exfoliated from a single crystal FeTe$_{\text{0.6}}$Se$_{\text{0.4}}$ onto a SiO$_{2}$/Si substrate. This measurement spans the wavelength range of 450 nm to 1100 nm and temperatures from 4 K to 295 K, with the optical fields aligned in-plane with the sample plane. We obtained the complex refractive index $\tilde{n}$ by employing the two-Drude band model for the dielectric function and use the transfer matrix method~\cite{MacleodBook} to fit thickness-dependent absolute reflection spectra from Fe(Te,Se)/SiO$_{2}$/Si stack. At 4 K, the refractive index and the extinction coefficient vary in the range of 2.81 - 4.26 and 2.24 - 3.21, respectively. In the normal state (295 K), they vary between 2.79 - 4.38 and 2.14 - 3.04. The two-Drude model reveals a stronger Drude component $\omega_{p,1} \simeq $ 27763  cm$^{-1}$ and the scattering rate $\gamma_{1}  \simeq$ 2521 cm$^{-1}$ at 4K. The other Drude component is $\omega_{p,2} \simeq $ 3184  cm$^{-1}$, with a weaker scattering rate $\gamma_{2}  \simeq$ 93 cm$^{-1}$ at 4K, which increases to $\simeq$ 950 cm$^{-1}$  at 295 K. 

\vspace{2mm}
\noindent \textbf{II. RESULTS AND DISCUSSION}

\vspace{2mm}
Figure 1 shows the experimental configuration of the optical reflectance measurement. A nanometer-thin flake of Fe(Te,Se) was exfoliated from the bulk single crystal FeTe$_{\text{0.6}}$Se$_{\text{0.4}}$ using the conventional scotch-tape method \cite{Huang2015} and transferred to a  SiO$_2$/Si substrate. An optical image of the flake under study is shown in Fig. 1(inset, lower left panel (b)), where the color variation across the flake depends both on the number of layers and on the thickness of SiO$_{2}$ layer. Here, the thickness of SiO$_{2}$ is 278 nm. The crystal structure of Fe(Te,Se), as shown in Fig. 1(inset, right panel (a)), is composed of a stack of Fe, Te/Se layers where a square lattice of Fe atoms is sandwiched between two Se/Te layers. The interlayer coupling is crucial for the electronic and magnetic properties of Fe(Te,Se). The lattice constant along the $c$-axis is 6.1 $\text{\AA}$, while 3.8 $\text{\AA}$ in the $a$-$b$ plane. For the reflectance measurements, the sample was illuminated using broadband emission from a tungsten halogen light source, which was then focused onto the sample using an objective (NA = 0.8) with a spot size of $\sim$3 $\mu$m. The same objective collected the reflected light and directed it to a spectrometer with a spectral resolution of 1 nm using a beam splitter. Only the \textit{in-plane} refractive index is measured in our experimental configuration as the optical fields lie in the sample plane. To determine the thickness of the exfoliated flake, we employed atomic force microscopy (AFM). Figure 1 (inset, upper left panel (c)) shows an AFM image of the Fe(Te,Se) flake under study and the height profile across the vertical line. From the height profile, we extracted the height of the flake as 7.6 nm, corresponding to $\sim 12$-layers.

We performed room temperature Raman spectroscopy to identify Fe(Te,Se) flakes using a confocal inVia Raman microscope with an excitation laser wavelength of 532 nm. A 50X objective microscope lens was utilized to focus the laser beam on the flake. The acquisition time was set to 5 seconds to achieve well-resolved peaks. We probed different sample locations to ensure the spectra's repeatability. Figure 2(a) shows the room-temperature Raman spectra in low wavenumber range 100 - 550 cm$^{-1}$ for 7.6 nm flake. In the range 100 - 250 cm$^{-1}$, this thick flake exhibits two peaks P$_{1,2}$ at 155.6 $\pm$ 0.1 cm$^{-1}$ and 203.8 $\pm$ 0.7 cm$^{-1}$ with a linewdith of 34.8$\pm$ 0.5 cm$^{-1}$ and 20.1 $\pm$ 2.6 cm$^{-1}$, respectively. These peaks P$_{1,2}$ have been attributed to the A$_{1g}$ mode associated with the out-of-plane vibration for Te/Se and the B$_{1g}$ mode for the Fe atom vibrations along the $c$-axis, respectively. Okazaki \textit{et al.} observed  A$_{1g}$ and B$_{1g}$ modes of FeTe$_{\text{0.6}}$Se$_{\text{0.4}}$ at 161 cm$^{-1}$ and 202 cm$^{-1}$, respectively~\cite{Okazaki2011}. However, large discrepancies have been observed in the literature \cite{Xia2009, Lopes2012, Lodhi2017} regarding the exact positions of these peaks.  For instance, Lopes \textit{et al.} reported similar Raman shifts at 120 $\pm$ 1 cm$^{-1}$ and 139 $\pm$ 1 cm$^{-1}$ in FeTe$_{\text{0.5}}$Se$_{\text{0.5}}$ sample \cite{Lopes2012}. 

Figure 2(b) shows the Raman spectra from 7.6 nm and 17.0 nm thick regions of the same flake under identical excitation power, integration time, and temperature conditions. As the thickness of the flakes increases to 17.0 nm, the position of A$_{1g}$ mode does not show any variation; however, the position of B$_{1g}$ mode is red-shifted to 195.6 $\pm$ 0.2 cm$^{-1}$. Furthermore, the linewidth of B$_{1g}$ mode reduces by more than two-fold for the 17.0 nm thick flake compared to the 7.6 nm flake. The linewidth of the Raman modes is associated with the lifetime of the phonon modes. A broader linewidth in thicker samples indicates a shorter lifetime due to increased scattering from grain boundaries, defects, and interlayer interactions. The spectral feature at $\sim$ 300 cm$^{-1}$ and $\sim$ 520 cm$^{-1}$ in Fig. 2(a,b) are associated with silicon substrate. 

Figure 3(a,d) shows the absolute reflectance spectra of our heterostructure stack Fe(Te,Se)/SiO$_2$/Si at 4 K and 295 K, respectively. The value of absolute reflectance reaches 0.4 for a wider range from 700 nm to 900 nm; with a strong absorption dip observed from 500 nm to 600 nm, which might be attributed to a strong absorption in this range. Next, to extract the complex refractive index $\tilde{n} = \sqrt{\tilde{\epsilon}}$, we employed the two-Drude model to describe the complex dielectric function of Fe(Te,Se)~\cite{Homes2011, Pimenov2013, Perucchi2014, Homes2015} and used transfer matrix method~\cite{MacleodBook} to fit thickness-dependent absolute reflection spectra. In the two-Drude model, the complex dielectric function is described by the standard Drude-Lorentz oscillator model \cite{DresselBook} with two Drude terms as follows:
\begin{equation}
\tilde{\epsilon}(\omega) = \epsilon_{\infty} - \frac{\omega^{2}_{p,1}}{\omega^{2}+i\omega\gamma_{1}}-\frac{\omega^{2}_{p,2}}{\omega^{2}+i\omega\gamma_{2}}+\sum_{j}\frac{\Omega^{2}_{j}}{\omega^{2}_{j}-\omega^{2}-i\omega\Gamma_{j}}
\end{equation}
where $\epsilon_{\infty}$ is the real part at high frequency, $\omega_{p,1(2)}$ and $\Gamma_{1(2)}$ are the plasma frequency and scattering rates for the Drude carriers, respectively. Contributions from the $j^{th}$-bound excitations is quantified by the parameters $\Omega_{j}$ (strength), $\omega_{j}$ (position), and $\gamma_{j}$ (width). We used the dielectric function of Eq. (1) with two Lorentzian terms to fit the measured absolute reflectance data by varying the model parameters to match the experimental data. 

Figure 3(a,d) shows the experimental data (black line) and the fitting curve (red line). From the fitting, we obtained the parameters of two Drude components $(\omega_{p,1/2},\gamma_{1/2})$ as:  $\omega_{p,1} \simeq $ 27736  cm$^{-1}$, $\gamma_{1}  \simeq$ 2512 cm$^{-1}$ and $\omega_{p,2} \simeq $ 3184  cm$^{-1}$, $\gamma_{2}  \simeq$ 93 cm$^{-1}$ at 4 K. At 295 K, those parameters become $\omega_{p,1} \simeq $ 26672  cm$^{-1}$, $\gamma_{1}  \simeq$ 3752 cm$^{-1}$ and $\omega_{p,2} \simeq $ 1000  cm$^{-1}$, $\gamma_{2}  \simeq$ 950 cm$^{-1}$. These values align with those reported for sister compounds FeTe$_{\textit{x}}$Se$_{1-\textit{x}}$~\cite{Homes2011, Pimenov2013, Perucchi2014, Homes2015} using the two-Drude model, where their multi-band nature reveals a stronger Drude component $(\omega_{p,1}, \gamma_{1})$ that exhibit some temperature-dependence and a weaker Drude component $(\omega_{p,2},\gamma_{2})$ with strongly temperature-dependent scattering rate (See supporting information Fig. S8). Furthermore, the broad Drude band $(\omega_{p,1},\gamma_{1})$ of FeTe$_{\text{0.6}}$Se$_{\text{0.4}}$ dominates the spectral weight ($\omega_{p,1}^{2}/\omega_{p,2}^{2}\gg$1) when compared to the narrower band $(\omega_{p,2},\gamma_{2})$. However,  the overall plasma frequency $\omega_{p} = (\omega^{2}_{p,1}+\omega^{2}_{p,2})^{1/2} \approx$ 27918 cm$^{-1}$ is higher, which may result from the difference in their conductivity.

Figures 3(b-f) show the refractive index and extinction coefficient of Fe(Te,Se) for photon wavelengths from 450 to 1100 nm at 4 K and 295 K. We observed both normal ($dn/d\lambda <0$) and abnormal ($dn/d\lambda >0$) dispersion in Fe(Te,Se) within specific spectral range. The refractive index and the extinction coefficient vary in the range of 2.81 - 4.26 and 2.24 - 3.21, respectively, at 4 K. At 295 K, they vary in the range of 2.79 - 4.38 and 2.14 - 3.04. In particular, at 800 nm, we have $k$ = 2.97 at 295 K. On the other hand, for the parent compound FeSe on CaF$_{2}$ substrate $k$ = 2.33 \cite{Gerber}. A higher extinction coefficient of Fe(Te,Se) is advantageous for photodetection applications~\cite{Natarajan2012}. To compare other high-$T_c$ superconductors with Fe(Te,Se), we have studied optical constants for Bi$_2$Sr$_2$CaCu$_2$O$_{8+\delta}$ (BSCCO) at 4 K (See supporting information Fig S8). In the entire spectral range (450 -1100 nm), the extinction coefficient for Fe(Te,Se) is higher than BSCCO, thus making it a promising candidate for high-temperature single-photon detectors~\cite{Natarajan2012, Merino2023, Charaev2023}.

Next, we studied the temperature-dependent reflection spectra of the flake from 4 K to 295 K for photon wavelengths from 450 to 1100 nm. The plot for differential reflectivity $ \Delta(\mathcal{R})$ of Fe(Te,Se) defined as 
\begin{equation}
    \Delta(\mathcal{R})=\frac{\mathcal{R}_{\text{Fe(Te,Se)}}-\mathcal{R}_{\text{SiO}_{\text{2}}/\text{Si}}}{\mathcal{R}_{\text{SiO}_{\text{2}}/\text{Si}}}
\end{equation}
is presented in Supporting information (See section S3). We employed the transfer matrix approach with the dielectric function of Fe(Te,Se) given by Eq.(1) to extract temperature-dependent ($n,k$), and the plasma frequencies and the scattering rates for two Drude components (See Section S4). We observed a sharp decrease in the scattering rate $\gamma_{1}$ at low temperatures ($<T_{c}$). Similar to the previous studies~\cite{Homes2011,Homes2015}, we observed an increase in the decay rate $\gamma_{2}$ with temperature that  has a quadratic form. The parameters corresponding to the two Lorentzian terms ($\omega_{1/2}, \Omega_{1/2}, \Gamma_{1/2}$) obtained from fitting the reflection data are summarized in section S8 of the Supporting information. 

Figure 4(a,b) plots the temperature-dependent refractive index and extinction coefficient from 4 K to 295 K, with a 0.2 offset on the $y$-axis for visual clarity for the 10 - 295 K temperature range. With increasing temperature, we have observed a blue shift in oscillator position at low energy ($\approx{450}$ nm). For instance, a blue shift of $\approx$ 11 nm in the position of the refractive index peak at 10 K ($\lambda \approx$ 482 nm) compared to 4 K ($\lambda \approx $ 493 nm) and the corresponding change $\Delta n = n_{10\,\text{K}}- n_{4\,\text{K}}\approx 0.011$. Similarly, a blue shift of $\approx$ 11 nm in the position of the extinction coefficient minima at 10 K ($\lambda \approx$ 542 nm) compared to 4 K ($\lambda \approx $ 553 nm) and the corresponding change $\Delta k = k_{10\,\text{K}}- k_{4\,\text{K}}\approx 0.067$. The variation of plasma frequencies $\omega_{p,1}$ and $\omega_{p,2}$ and the scattering rates $\Gamma_{1}$ and $\Gamma_{2}$ has been presented in Supporting Information S4. We observed that the plasma frequency and scattering terms for both the broad and narrow components show little to no variation up to 70 K for Fe (Te,Se) apart from a strong narrow component $\omega_{p,2}$ 4 K for Fe(Te,Se). On the other hand, the scattering rate for the narrow component $\Gamma_{2}$ gradually increases with temperature. At 295 K, it reaches about a quarter of the value of $\Gamma_{2}$ and is an order of magnitude greater than its value at 4 K. 

\vspace{2mm}
\noindent \textbf{III. CONCLUSIONS}

\vspace{2mm}
In summary, we report the measurement of the complex-valued in-plane refractive index $\tilde{n} = n+ik$ of nanometer-thin flakes of multiband superconductor Fe(Te,Se) over the 450 to 1100 nm wavelength range from 4 K to 295 K. We obtained the complex refractive index $\tilde{n}$ by employing the two-Drude band model and the transfer matrix method to fit thickness-dependent reflection spectra from our heterostructure stack Fe(Te,Se)/SiO$_2$/Si. In room-temperature Raman spectra, we observed two peaks P$_{1,2}$ at 155.6 $\pm$ 0.1 cm$^{-1}$ and 203.8 $\pm$ 0.7 cm$^{-1}$ with a linewdith of 34.8$\pm$ 0.5 cm$^{-1}$ and 20.1 $\pm$ 2.6 cm$^{-1}$, respectively. These peaks P$_{1,2}$ have been attributed to the A$_{1g}$ mode for Te/Se and the B$_{1g}$ mode for the Fe atom vibrations, respectively. For a thicker (17.0 nm) flake, we observed B$_{1g}$ mode is red-shifted by 8.2 $\pm$ 0.7 cm$^{-1}$ but no noticeable change in the position of A$_{1g}$ mode compared to 7.6 nm Fe(Te,Se) flake. In the superconducting state (4 K), the refractive index and the extinction coefficient vary in the range of 2.81 - 4.26 and 2.24 - 3.21, respectively. In the normal state (295 K), they vary between 2.79 - 4.38 and 2.14 - 3.04. Our temperature-dependent reflection spectra yield a stronger and broader Drude response $\omega_{p,1} \simeq $ 27736  cm$^{-1}$, with a significant scattering rate $\gamma_{1}  \simeq$ 2512 cm$^{-1}$ that is essentially independent of temperature at the low-temperature regime. The other Drude response $\omega_{p,2} \simeq $ 3184  cm$^{-1}$, with weaker scattering rate $\gamma_{2}  \simeq$ 93 cm$^{-1}$ which increases by an order of magnitude to $\simeq$ 950 cm$^{-1}$. The high extinction coefficient of FeTe$_{\text{0.6}}$Se$_{\text{0.4}}$ in the visible and near-IR regions of the electromagnetic spectrum, combined with its high-$T_c$ superconductivity, makes it an exciting material for a high-temperature single-photon detector~\cite{Natarajan2012, Merino2023, Charaev2023}. Furthermore, the 2D nature of Fe(Te,Se) superconductor can be leveraged to create heterostructures with other 2D materials, such as hexagonal boron nitride (\textit{h}BN)~\cite{Akbari2021,Zhou2024,Jha2021,Akbari2022}, and control emission characteristics (emission rates, radiation pattern) of color centers in \textit{h}BN.

\vspace{3mm}
\noindent \textbf{SUPPLEMENTARY MATERIAL}

\vspace{2mm}
Experimental setup, AFM Measurement for 17.0 nm Fe(Te,Se) flake, differential reflectivity for 7.6 nm thin Fe(Te,Se) flake, temperature-dependence of Drude parameters for 7.6 nm Fe(Te,Se) flake, absolute reflectance for 17.0 nm and 7.6 nm Fe(Te,Se) flake, complex refractive index for 17.0 nm Fe(Te,Se) flake at 4 K and 295 K, absolute reflectance and coefficients of complex refractive index for BSCCO, fitting parameters (Lorentz-oscillators) for 7.6 nm flake. 

\vspace{3mm}
\noindent \textbf{ACKNOWLEDGMENTS}

\vspace{2mm}
This work was supported by the Syracuse University Start-up Funds. 

\vspace{3mm}
\noindent \textbf{DATA AVAILABILITY}

\vspace{2mm}
The data that support the findings of this study are available from the corresponding author upon reasonable request.

\newpage
\begin{figure}[t!]
	{\includegraphics[scale = 1.5]{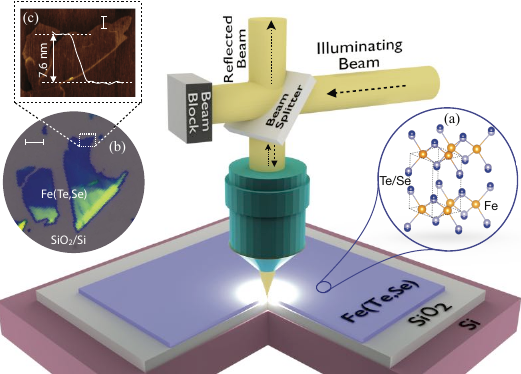}}
  \caption{Experimental schematic and thickness characterization of Fe(Te,Se) flake. A nanometer-thin flake was exfoliated from a single crystal FeTe$_{\text{0.6}}$Se$_{\text{0.4}}$ and transferred onto SiO$_2$/Si substrate using the conventional scotch-tape method. The thickness of the SiO$_2$ layer is 278 nm. For our reflectance measurements, the same objective lens was used for broadband illumination and collection of the reflected light over the entire temperature range of 4 - 295 K. The lower right inset panel (a) shows the crystal structure of Fe(Te,Se) space group P4/nmm (No. 129), with room-temperature lattice parameters of $a$ = $b \approx$ 3.8 $\text{\AA},c \approx$ 6.1 $\text{\AA}$. The lower left inset panel (b) shows the optical image of the flake of varying thicknesses used in this study. The upper left inset panel (c) shows an atomic force microscopy topography mapping of the flake. The vertical line trace gives a thickness of 7.6 nm which corresponds to $\sim$ 12 layers of the boxed region highlighted in (b).}
 \end{figure}
 \newpage
\begin{figure}[t!]
	{\includegraphics[scale = 1.0]{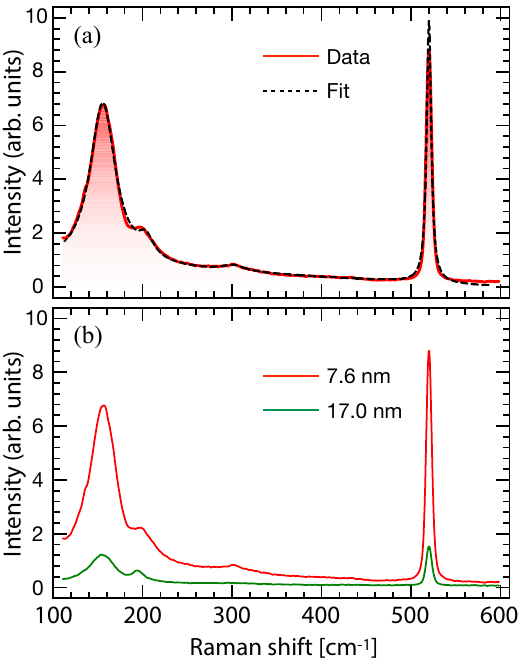}}
  \caption{Room-temperature Raman spectroscopy of Fe(Te,Se) flakes. (a) Raman spectra for the 7.6 nm thin flake with multiple Lorentzian curve fitting. In the range 100 - 250 cm$^{-1}$, the sample exhibits two peaks at 155.6 $\pm$ 0.1 cm$^{-1}$ and 203.8 $\pm$ 0.7 cm$^{-1}$ with a linewidth of 34.8 $\pm$ 0.5 cm$^{-1}$ and 20.1 $\pm$ 2.6 cm$^{-1}$ respectively. (b) Comparison of Raman spectra of 7.6 nm thin flake with 17.0 nm thin flake. With a thicker sample, we observed B$_{1g}$ mode is red-shifted by 8.2 $\pm$ 0.7 cm$^{-1}$ but no noticeable change in the position of A$_{1g}$ mode. The spectral feature at $\sim$ 300 cm$^{-1}$ and $\sim$ 520 cm$^{-1}$ in Fig. 2(a,b) are associated with silicon substrate.}
 \end{figure}
  \newpage
\begin{figure}[t!]
	{\includegraphics[scale = 0.90]{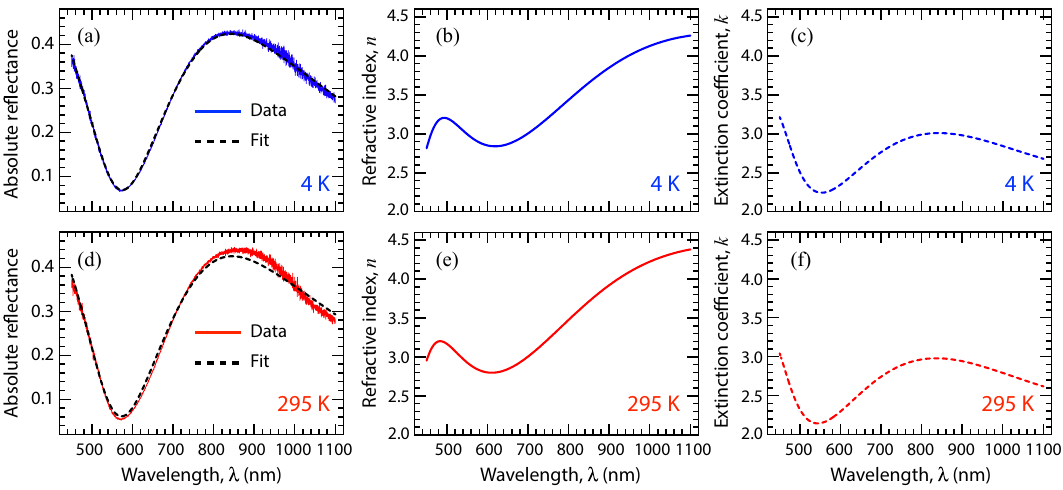}}
  \caption{Optical constants of the Fe(Te,Se) flake at 4 K and 295 K. The plot of the absolute reflectance and fitted curve of the 7.6 nm thick Fe(Te,Se) flake at (a) 4K and (d) 295 K, respectively. The absolute reflectance was obtained by normalizing the sample's reflection spectra with the reflection spectra of a silver mirror. We fitted the measured reflectance spectra of our heterostructure stack Fe(Te,Se)/SiO$_2$/Si using the transfer matrix method with the dielectric constant of Fe(Te,Se) given by Eq.(1). Plot of the calculated refractive index (\textit{n}) and the extinction coefficient (\textit{k}) of the Fe(Te,Se) flake at (b,c) 4K and (d,e) 295 K. Extinction coefficient minima, within our measurement window, at 295 K is 2.14 compared to 2.24 at 4 K. Similarly at 1100 nm the refractive index at 295 K is 4.38 which is higher than 4.26 at 4 K.}
 \end{figure}
\newpage
\begin{figure}[t!]
	{\includegraphics[scale = 0.90]{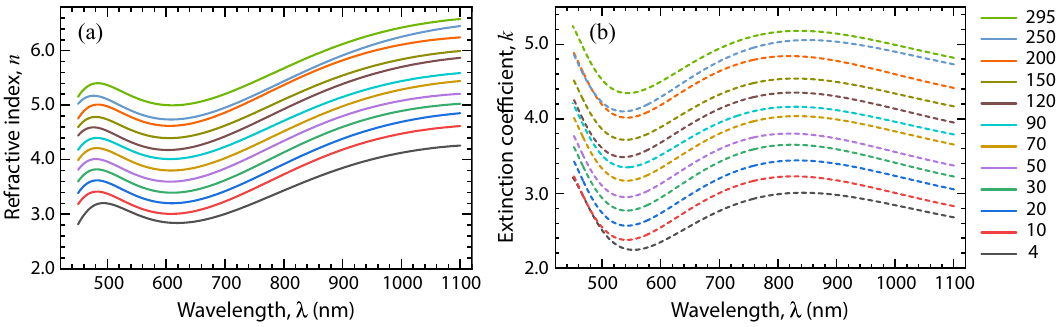}}
  \caption{Temperature-dependent in-plane complex refractive index of Fe(Te,Se) flake. Plot of the (a) refractive index $n$ and the (b) extinction coefficient $k$ of the 7.6 nm thin  Fe(Te,Se) flake against the wavelength at various temperatures. We observed a blue shift of $\approx$ 11 nm in the position of the refractive index peak at 10 K ($\lambda \approx$ 482 nm) compared to 4 K ($\lambda \approx $ 493 nm). Similarly, a blue shift of $\approx$ 11 nm in the position of the extinction coefficient minima at 10 K ($\lambda \approx$ 542 nm) compared to 4 K ($\lambda \approx $ 553 nm). In (a,b), the curves corresponding to the temperature range 10 - 295 K have been shifted vertically by 0.2 units for visual clarity.}
 \end{figure}


\newpage
\begin{thebibliography}{99}
\bibitem{Kamihara2008} Y. Kamihara, T. Watanabe, M. Hirano, and H. Hosono, “Iron-Based Layered Superconductor La[O$_{1-x}$F$_{x}$]FeAs (x=0.05 - 0.12) with T$_{c} =$ 26 K,” J. Am. Chem. Soc. 130(11), 3296–3297 (2008).
\bibitem{Muller1986} J. G. Bednorz, and K. A. Müller, “Possible high T$_c$ superconductivity in the barium-lanthanum-copper-oxygen system,” Z. Physik B - Condensed Matter 64(2), 189–193 (1986).
\bibitem{Chen2008}G. F. Chen, Z. Li, D. Wu, G. Li, W. Z. Hu, J. Dong, P. Zheng, J. L. Luo, and N. L. Wang, “Superconductivity at 41 K and Its Competition with Spin-Density-Wave Instability in Layered ${\mathrm{CeO}}_{1\ensuremath{-}x}{\mathrm{F}}_{x}\mathrm{FeAs}$,” Phys. Rev. Lett. 100(24), 247002 (2008).
\bibitem{Ren2008b} Z. A. Ren, J. Yang, W. Lu, W. Yi, X. L. Shen, Z. C. Li, G. C. Che, X. L. Dong, L. L. Sun, F. Zhou, and Z. X. Zhao, “Superconductivity in the iron-based F-doped layered quaternary compound Nd[O$_{1-x}$F$_{x}$]FeAs,” EPL 82(5), 57002 (2008).
\bibitem{Ren2008a} Z. A. Ren, J. Yang, W. Lu, W. Yi, G. C. Che, X. L. Dong, L. L. Sun, and Z.X. Zhao, “Superconductivity at 52 K in iron based F doped layered quaternary compound Pr[O$_{1-x}$F$_{x}$]FeAs,” Materials Research Innovations 12(3), 105–106 (2008).
\bibitem{Ren2008c} R. Zhi-An, L. Wei, Y. Jie, Y. Wei, S. Xiao-Li, Zheng-Cai, C. Guang-Can, D. Xiao-Li, S. Li-Ling, Z. Fang, and Z. Zhong-Xian, “Superconductivity at 55 K in Iron-Based F-Doped Layered Quaternary Compound  Sm[O$_{1-x}$F$_{x}$]FeAs,” Chinese Phys. Lett. 25(6), 2215 (2008).
\bibitem{Hirschfeld2011} P. J. Hirschfeld, M. M. Korshunov, and I. I. Mazin, “Gap symmetry and structure of Fe-based superconductors,” Rep. Prog. Phys. 74(12), 124508 (2011).
\bibitem{Scalapino1995} D. J. Scalapino, “The case for $d_{x^{2}-y^{2}}$ pairing in the cuprate superconductors.,” Physics Reports 250(6), 329–365 (1995).
\bibitem{Hsu2008} F. C. Hsu, J. Y. Luo, K. W. Yeh, T. K. Chen, T. W. Huang, P. M. Wu, Y. C. Lee, Y. L. Huang, Y. Y. Chu, D. C. Yan, and M. K. Wu, “Superconductivity in the PbO-type structure $\alpha$-FeSe,” Proceedings of the National Academy of Sciences 105(38), 14262–14264 (2008).
\bibitem{Yeh2008a} K.-W. Yeh, T.-W. Huang, Y. Huang, T.-K. Chen, F.-C. Hsu, P.M. Wu, Y.-C. Lee, Y.-Y. Chu, C.-L. Chen, J.-Y. Luo, D.-C. Yan, and M.-K. Wu, “Tellurium substitution effect on superconductivity of the $\alpha$-phase iron selenide,” Europhys. Lett. 84, 37002 (2008).
\bibitem{Fang2008} M. H. Fang, H.M. Pham, B. Qian, T.J. Liu, E.K. Vehstedt, Y. Liu, L. Spinu, and Z.Q. Mao, “Superconductivity close to magnetic instability in FeTe$_{x}$(Se$_{1-x})_{0.82}$,” Phys. Rev. B 78(22), 224503 (2008).
\bibitem{Yeh2008b} K. W. Yeh, H. C. Hsu, T. W. Huang, P. M. Wu, Y. L. Huang, T. K. Chen, J. Y. Luo, and M. K. Wu, “Se and Te Doping Study of the FeSe Superconductors,” J. Phys. Soc. Jpn. 77(Suppl.C), 19–22 (2008).
\bibitem{Paglione2010} J. Paglione, and R.L. Greene, “High-temperature superconductivity in iron-based materials,” Nature Phys 6(9), 645–658 (2010).
\bibitem{Rameau2019} J. D. Rameau, N. Zaki, G. D. Gu, P. D. Johnson, and M. Weinert, “Interplay of paramagnetism and topology in the Fe-chalcogenide high-${T}_{c}$ superconductors,” Phys. Rev. B 99(20), 205117 (2019).
\bibitem{McLaughlin2021} N. J. McLaughlin, H. Wang, M. Huang, E. Lee-Wong, L. Hu, H. Lu, G. Q. Yan, G. Gu, C. Wu, Y-Z You, C. R. Du, "Strong Correlation Between Superconductivity and Ferromagnetism in an Fe-Chalcogenide Superconductor," Nano Letters 21, 7277 (2021).
\bibitem{Zhang2018} P. Zhang, K. Yaji, T. Hashimoto, Y. Ota, T. Kondo, K. Okazaki, Z. Wang, J. Wen, G.D. Gu, H. Ding, and S. Shin, "Observation of topological superconductivity on the surface of an iron-based superconductor," Science 360, 182 (2018).
\bibitem{Zhang2019} P. Zhang, Z. Wang, X. Wu, K. Yaji, Y. Ishida, Y. Kohama, G. Dai, Y. Sun, C. Bareille, K. Kuroda, T. Kondo, K. Okazaki, K. Kindo, X. Wang, C. Jin, J. Hu, R. Thomale, K. Sumida, S. Wu, K. Miyamoto, T. Okuda, H. Ding, G.D. Gu, T. Tamegai, T. Kawakami, M. Sato, and S. Shin, “Multiple topological states in iron-based superconductors,” Nature Phys 15(1), 41–47 (2019).
\bibitem{Wang2018} D. Wang, L. Kong, P. Fan, H. Chen, S. Zhu, W. Liu, L. Cao, Y. Sun, S. Du, J. Schneeloch, R. Zhong, G. Gu, L. Fu, H. Ding, and H.-J. Gao, “Evidence for Majorana bound states in an iron-based superconductor,” Science 362, 333 (2018).
\bibitem{Kong2019} L. Kong, S. Zhu, M. Papaj, H. Chen, L. Cao, H. Isobe, Y. Xing, W. Liu, D. Wang, P. Fan, Y. Sun, S. Du, J. Schneeloch, R. Zhong, G. Gu, L. Fu, H.-J. Gao, and H. Ding, “Half-integer level shift of vortex bound states in an iron-based superconductor,” Nat. Phys. 15(11), 1181–1187 (2019).
\bibitem{Machida2019} T. Machida, Y. Sun, S. Pyon, S. Takeda, Y. Kohsaka, T. Hanaguri, T. Sasagawa, and T. Tamegai, "Zero-energy vortex bound state in the superconducting topological surface state of Fe(Se,Te)," Nat. Mater. 18, 811 (2019).
\bibitem{Li2015} F. Li, H. Ding, C. Tang, J. Peng, Q. Zhang, W. Zhang, G. Zhou, D. Zhang, C.-L. Song, K. He, S. Ji, X. Chen, L. Gu, L. Wang, X.-C. Ma, and Q.-K. Xue, “Interface-enhanced high-temperature superconductivity in single-unit-cell FeTe$_{x}$Se$_{1-x}$ films on SrTiO$_{3}$.,” Phys. Rev. B 91(22), 220503 (2015).
\bibitem{Sales2009} B.C. Sales, A.S. Sefat, M.A. McGuire, R.Y. Jin, D. Mandrus, and Y. Mozharivskyj, “Bulk superconductivity at 14 K in single crystals of Fe$_{1-y}$Te$_{x}$Se$_{1-x}$,” Phys. Rev. B 79(9), 094521 (2009).
\bibitem{Pourret2011} A. Pourret, L. Malone, A.B. Antunes, C.S. Yadav, P.L. Paulose, B. Fauqué, and K. Behnia, “Strong correlation and low carrier density in Fe$_{1-y}$Te$_{0.6}$Se$_{0.4}$ as seen from its thermoelectric response,” Phys. Rev. B 83(2), 020504 (2011).
\bibitem{Lanata2013} N. Lanatà, H.U.R. Strand, G. Giovannetti, B. Hellsing, L. de’ Medici, and M. Capone, “Orbital selectivity in Hund’s metals: The iron chalcogenides,” Phys. Rev. B 87(4), 045122 (2013).
\bibitem{X2} A. Subedi, L. Zhang, D. J. Singh, and M. H. Du, “Density functional study of FeS, FeSe, and FeTe: Electronic structure, magnetism, phonons, and superconductivity,” Phys. Rev. B 78(13), 134514 (2008).
\bibitem{X3} F. Chen, B. Zhou, Y. Zhang, J. Wei, H. W. Ou, J.-F. Zhao, C. He, Q.-Q. Ge, M. Arita, K. Shimada, H. Namatame, M. Taniguchi, Z.-Y. Lu, J. Hu, X.-Y. Cui, and D.L. Feng, “Electronic structure of ${\text{Fe}}_{1.04}{\text{Te}}_{0.66}{\text{Se}}_{0.34}$,” Phys. Rev. B 81(1), 014526 (2010).
\bibitem{X4} A. Tamai, A.Y. Ganin, E. Rozbicki, J. Bacsa, W. Meevasana, P.D.C. King, M. Caffio, R. Schaub, S. Margadonna, K. Prassides, M.J. Rosseinsky, and F. Baumberger, “Strong Electron Correlations in the Normal State of the Iron-Based ${\mathrm{FeSe}}_{0.42}{\mathrm{Te}}_{0.58}$ Superconductor Observed by Angle-Resolved Photoemission Spectroscopy,” Phys. Rev. Lett. 104(9), 097002 (2010).
\bibitem{X5} P. Zhang, P. Richard, N. Xu, Y.-M. Xu, J. Ma, T. Qian, A.V. Fedorov, J.D. Denlinger, G.D. Gu, and H. Ding, “Observation of an electron band above the Fermi level in FeTe0.55Se0.45 from in-situ surface doping,” Applied Physics Letters 105(17), 172601 (2014).
\bibitem{R1} H. Miao, P. Richard, Y. Tanaka, K. Nakayama, T. Qian, K. Umezawa, T. Sato, Y.-M. Xu, Y.B. Shi, N. Xu, X.-P. Wang, P. Zhang, H.-B. Yang, Z.-J. Xu, J.S. Wen, G.-D. Gu, X. Dai, J.-P. Hu, T. Takahashi, and H. Ding, “Isotropic superconducting gaps with enhanced pairing on electron Fermi surfaces in FeTe$_{\text{0.55}}$Se$_{\text{0.45}}$,” Phys. Rev. B 85(9), 094506 (2012).
\bibitem{R2} K. Nakayama, R. Tsubono, G.N. Phan, F. Nabeshima, N. Shikama, T. Ishikawa, Y. Sakishita, S. Ideta, K. Tanaka, A. Maeda, T. Takahashi, and T. Sato, “Orbital mixing at the onset of high-temperature superconductivity in FeTe$_x$Se$_{1-x}$/CaF$_2$,” Phys. Rev. Res. 3(1), L012007 (2021).
\bibitem{Homes2011} C.C. Homes, A. Akrap, J. Wen, Z. Xu, Z. Wei Lin, Q. Li, and G. Gu, “Optical properties of the iron-chalcogenide superconductor FeTe0.55Se0.45,” Journal of Physics and Chemistry of Solids 72(5), 505–510 (2011).
\bibitem{Pimenov2013} A. Pimenov, S. Engelbrecht, A.M. Shuvaev, B.B. Jin, P.H. Wu, B. Xu, L.X. Cao, and E. Schachinger, “Terahertz conductivity in FeSe0.5Te0.5 superconducting films,” New J. Phys. 15(1), 013032 (2013).
\bibitem{Perucchi2014} A. Perucchi, B. Joseph, S. Caramazza, M. Autore, E. Bellingeri, S. Kawale, C. Ferdeghini, M. Putti, S. Lupi, and P. Dore, “Two-Band Conductivity of a FeSe$_{0.5}$Te$_{0.5}$ Film by Reflectance Measurements in the Terahertz and Infrared Range,” Supercond. Sci. Technol. 27(12), 125011 (2014).
\bibitem{Homes2015} C.C. Homes, Y.M. Dai, J.S. Wen, Z.J. Xu, and G.D. Gu, “${\mathrm{FeTe}}_{0.55}{\mathrm{Se}}_{0.45}$: A multiband superconductor in the clean and dirty limit,” Phys. Rev. B 91(14), 144503 (2015).
\bibitem{Li2017} C. Li, S. Sun, S. Wang, H. Lei, "Enhanced superconductivity and anisotropy of FeTe$_{0.6}$Se$_{0.4}$ single crystals with Li-NH3 intercalation," Phys. Rev. B 96, 134503 (2017).
\bibitem{Lin2023} C. W Lin, I. N. Chen, Z. Lei, and L. M. Wang, "Two-dimensional-like superconducting properties and weak antilocalization transport in FeSe$_{0.4}$Te$_{0.6}$ single crystals: Topology-driven magnetotransport," Phys. Rev. B 108, 214509 (2023).
\bibitem{Miyazawa2021} T. Miyazawa, N. Tadokoro, S. Horikawa, T. Tamegai, Y. Sun, and H. Kitano, "Focused ion beam microfabrication of single-crystal nanobridge toward Fe(Te, Se)-based Josephson device," J. Phys.: Conf. Ser. 1975 012010 (2021).
\bibitem{Qiu2023} G. Qiu, H. Yang, L. Hu, H. Zhang, C. Chen, Y. Lyu, C. Eckberg, P. Deng, S. Krylyuk,  A. V. Davydov, R. Zhang, and  K. L. Wang, "Emergent ferromagnetism with superconductivity in Fe(Te,Se) van der Waals Josephson junctions," Nat. Commun 14, 1 (2023).
\bibitem{Ohnishi2021} K. Ohnishi, S. Gupta, S. Kasahara, Y. Kasahara,  Y. Matsuda, E. Shigematsu,  R. Ohshima, Y.  Ando, and M. Shiraishi, M, "Observation of a superconducting state of a topological superconductor candidate, FeTe$_{0. 6}$Se$_{0. 4}$, equipping ferromagnetic electrodes with perpendicular magnetic anisotropy," Appl. Phys. Express 14, 093002 (2021).
\bibitem{Tanaka2017} Y. Tanaka, Y. Mizuguchi, and O. Miura, "FeTe$_{\text{0.6}}$Se$_{\text{0.4}}$ bulk single crystals with high critical current densities under magnetic fields," IOP Conf. Series: Journal of Physics: Conf. Series 871, 012064 (2017).
\bibitem{Rosfjord2006} K. M. Rosfjord, J. K. W. Yang, E. A. Dauler, A. J. Kerman, V. Anant, B. M. Voronov, G. N. Gol’tsman, and K. K. Berggren,"Nanowire Single-photon detector with an integrated optical cavity and anti-reflection coating," Opt. Express 14, 527 (2006). 
\bibitem{MacleodBook} H.A. Macleod, Thin-Film Optical Filters, Fourth Edition (Taylor Francis, 2010).
\bibitem{Huang2015} Y. Huang, E. Sutter, N.N. Shi, J. Zheng, T. Yang, D. Englund, H.-J. Gao, and P. Sutter, “Reliable Exfoliation of Large-Area High-Quality Flakes of Graphene and Other Two-Dimensional Materials,” ACS Nano 9(11), 10612–10620 (2015).
\bibitem{Okazaki2011} K. Okazaki, S. Sugai, S. Niitaka, and H. Takagi, “Phonon, two-magnon, and electronic Raman scattering of Fe${}_{1+y}$Te${}_{1\ensuremath{-}x}$Se${}_{x}$,” Phys. Rev. B 83(3), 035103 (2011).
\bibitem{Lopes2012} C. S. Lopes, C. E. Foerster, F. C. Serbena, P. R. Júnior, A. R. Jurelo, J. L. P. Júnior, P. Pureur, and A. L. Chinelatto, “Raman spectroscopy of highly oriented FeSe0.5Te0.5 superconductor,” Supercond. Sci. Technol. 25(2), 025014 (2012).
\bibitem{Xia2009}  T. L. Xia, D. Hou, S.C. Zhao, A.M. Zhang, G.F. Chen, J.L. Luo, N.L. Wang, J.H. Wei, Z.-Y. Lu, and Q.M. Zhang, “Raman phonons of $\ensuremath{\alpha}\text{-FeTe}$ and ${\text{Fe}}_{1.03}{\text{Se}}_{0.3}{\text{Te}}_{0.7}$ single crystals,” Phys. Rev. B 79(14), 140510 (2009).
\bibitem{Lodhi2017} P. D. Lodhi, V. P. S. Awana, and N. Kaurav, “Raman and X-Ray Diffraction Studies of Superconducting FeTe  $_{1- x}$  Se $_x$  Compounds,” J. Phys.: Conf. Ser. 836(1), 012046 (2017).
\bibitem{Kumar2010} P. Kumar, A. Kumar, S. Saha, D.V.S. Muthu, J. Prakash, S. Patnaik, U.V. Waghmare, A.K. Ganguli, and A.K. Sood, “Anomalous Raman Scattering from Phonons and Electrons of Superconducting FeSe$_0.82$,” Solid State Communications 150(13), 557–560 (2010).
\bibitem{DresselBook} M. Dressell and G. Grunner \textit{Electrodynamics of Solids} (Cambridge: Cambridge University Press 2002)
\bibitem{Gerber} S. Gerber, S.-L. Yang, D. Zhu, H. Soifer, J.A. Sobota, S. Rebec, J.J. Lee, T. Jia, B. Moritz, C. Jia, A. Gauthier, Y. Li, D. Leuenberger, Y. Zhang, L. Chaix, W. Li, H. Jang, J.-S. Lee, M. Yi, G.L. Dakovski, S. Song, J.M. Glownia, S. Nelson, K.W. Kim, Y.-D. Chuang, Z. Hussain, R.G. Moore, T.P. Devereaux, W.-S. Lee, P.S. Kirchmann, and Z.-X. Shen, “Femtosecond electron-phonon lock-in by photoemission and x-ray free-electron laser,” Science 357(6346), 71–75 (2017).
\bibitem{Natarajan2012} C. M. Natarajan, M. G. Tanner, and R. H. Hadfield, “Superconducting nanowire single-photon detectors: physics and applications,” Supercond. Sci. Technol. 25(6), 063001 (2012).
\bibitem{Merino2023} R. L. Merino, P. Seifert, J. D. Retamal, R. K. Mech, T. Taniguchi, K. Watanabe, K. Kadowaki, R. H Hadfield, D. K Efetov, “Two-dimensional cuprate nanodetector with single telecom photon sensitivity at T = 20 K," 2D Mater. 10, 021001 (2023).
\bibitem{Charaev2023} I. Charaev, D. A. Bandurin, A. T. Bollinger, I. Y. Phinney, I. Drozdov, M. Colangelo, B. A. Butters, T. Taniguchi, K. Watanabe, X. He, I. Bozovic, P. Jarillo-Herrero, K. K. Berggren, “Single-photon detection using high-temperature superconductors," Nature Nanotechnol 18, 343 (2023).
\bibitem{Akbari2021} H. Akbari, W. H Lin, B. Vest, P. K Jha, and H. A. Atwater, “Temperature-dependent spectral emission of hexagonal boron nitride quantum emitters on conductive and dielectric substrates,” Phys. Review Applied 15, 014036 (2021).
\bibitem{Zhou2024} J. Zhou, H. Lu, D. Chen, M. Huang, G. Q. Yan, F. Al-Matouq, J. Chang, D. Djugba, Z. Jiang, H. Wang, C. R. Du, “Sensing spin wave excitations by spin defects in few-layer-thick hexagonal boron nitride," Science Advances 10, eadk8495 (2024). 
\bibitem{Jha2021} P. K Jha, H. Akbari, Y. Kim, S. Biswas, and H. A . Atwater, “Nanoscale axial position and orientation measurement of hexagonal boron nitride quantum emitters using a tunable nanophotonic environment,” Nanotechnology 33, 015001 (2021).
\bibitem{Akbari2022} H. Akbari, S. Biswas, P. K. Jha, J. Wong, B. Vest, and H. A. Atwater, “Lifetime-limited and tunable quantum light emission in h-bn via electric field modulation,” Nano Letters 22, 7798 (2022).
\end{thebibliography}
\end{document}